# A Compact, Data-Logging Breath-Gas Analyzer

Shelby Lacouture, Mitchell Kelley, Noah Plues, Laszlo Hunyadi, Emily Sundman, Annette Sobel, and Robert V. Duncan, *Texas Tech University, Lubbock, Texas*

Respiratory ailments are increasing globally at an alarming rate and are currently one of the leading factors of death and infirmity worldwide. Among respiratory diseases, those linked to poor air quality and pollutants are increasing at a proportionally higher rate than those linked to viral or other factors. Diagnosing disorders of the respiratory system is often performed initially by routine physical examinations and questionnaires. Once most patients have symptoms that are severe enough to warrant clinical testing, the ailment could have already caused pulmonary damage. Clinical diagnosis involves the use of cumbersome, expensive equipment that measures different parameters separately, e.g., Capnography (CO2) and spirometry (bidirectional tidal mass flow). These disparate sets of data must then be interpreted collectively by a qualified medical practitioner. This paper details the design of a portable, inexpensive, mixed-signal data-logging system that measures a chosen set of parameters in exhaled breath from humans or animals. The data is a comprehensive set of pertinent gases and mass flow that when looked at simultaneously, gives a synergistic view of these interrelated breathing biomarkers and thus the state of the respiratory system as a whole. A mask-mounted, tabletop, and handheld version was developed for different applications. The system, when fully developed, would enable a new set of clinical vitals that only require a patient to breathe through a single, small device for a few moments. This new set of clinical vitals could enable the early diagnosis of many respiratory ailments, something that could have a large positive impact on disease prognosis and quality of life.

## I. INTRODUCTION

Worldwide, as of 2017, 544.9 million people (not accounting for undiagnosed or unreported cases) suffer from one form or another of chronic respiratory disease, an increase from 1990 of 39.8%, and chronic respiratory ailments are now a leading cause of death and disability worldwide [1]. Environmental factors appear to be playing much more of a role in respiratory ailments recently, as there has been a significant shift in the cause of the majority of these disorders from infectious diseases (e.g., tuberculosis, COVID-19, etc.) to those resulting from pollutants in the air (Asthma, COPD, etc.) [2]. The financial costs of breathing disorders are also enormous: in the US alone, medical claims resulting from asthma reached $7 billion ($5 billion for COPD) yearly [3]. When missed work, childcare and other factors are included, the costs for asthma alone have been estimated at $80 billion per year [4]. Despite these alarming statistics, research funding for respiratory diseases has historically lagged other health concerns [5]. The COVID-19 pandemic, however, has spawned a renewed interest in, and calls for, increased funding for medical research into all respiratory ailments [6]. Spirometry, coupled with patient questionnaires, is currently the most common approach to diagnosing many of the various respiratory illnesses [7]. These tests are often only performed after the onset of respiratory distress symptoms, however, and research indicates a need for early diagnosis to improve long-term outcomes, especially in children and young adults [8]. Respiratory ailments are certainly not limited to humans; cats, dogs, horses, and many other species suffer from the same breathing conditions people do [9]. The respiratory systems of many animals are very similar to humans, offering a comparative model for understanding the exact mechanisms,

progression of and possible treatments for humans [10]. The foregoing points out a serious need for improved and more widespread respiratory diagnostic testing and for more comprehensive research into respiratory ailments in humans and animals, and the effect of air quality on both. The research discussed in this document revolves around a portable system, named a Data Logging Sensor Integration Block (DLSIB). The system is an "intelligent" data logging unit that adds digital communication channels to that of a traditional analog data logger. The initial prototype unit currently measures (but is not limited to) Oxygen ($O_2$), Carbon Dioxide ($CO_2$), Nitric Oxide (NO), Mass flow, Temperature, and Relative Humidity (RH), of human or animal respiration. The unit can be (and has previously) been configured to perform pulse oximetry ($SpO_2$) in tandem as well. The system is also capable of measuring and recording surrounding local air quality (particulate matter, VOC's, HCHO, etc.) simultaneously.

**II. Mixed-Signal Datalogging System for Respiratory Biomarker Measurement**

The system created to measure the respiratory biomarkers discussed in the preceding sections is detailed below. The device is a mixed-signal (analog and digital) data logging system with specialized analog circuitry to read, and supply bias voltages (where necessary) to Electrochemical (EC) gas sensors. The system, comprising the main electronics, and the chosen sensors can be placed in a housing specifically meant for in-line readings between an individual and a CPAP machine or respirator unit, in a hand-held housing for clinical vitals, field deployment, or the consumer market, and finally in a housing designed specifically for veterinarian use, e.g., horses or livestock. The invention records (at a user chosen sample rate) various parameters of exhaled gas, including, but not limited to: Oxygen ($O_2$), Carbon Dioxide ($CO_2$), Nitric Oxide (NO), Mass flow, Temperature, and Relative Humidity (RH). Either analog or intelligent sensors can be added and removed at will. When used in stand-alone data logging mode, each recording is time and date stamped. The unit can make several recordings, limited only by file size (set sample rate and recording length) and available memory. Once data has been recorded in the non-volatile memory, it can be uploaded to a host computer for display and can be saved to a local drive for more detailed analysis. Data can also be logged real-time via a USB connection.

**A. DLSIB System**

Fig. 1 is a block diagram of the electrical components that comprise the core of the system, along with the sensors initially used for pulmonary functioning evaluation of equines. A microcontroller collects sensor data either in analog form from those sensors that produce a voltage or current output via the analog front end, or in digitized format from intelligent sensors using various digital communication protocols, currently: Serial Peripheral Interface (SPI), Inter-Integrated Circuit ($I^2C$) and Universal Asynchronous Receiver Transmitter (UART). The complement of sensors shown includes a mass flow sensor, an oxygen sensor, a carbon dioxide sensor, a Nitric Oxide sensor, and a combination temperature and RH sensor connected to the core system. The system has built-in analog processing, including specialized subsystems such as a Potentiostat circuit to read the raw output of 3-electrode EC sensors and Transimpedance amplifiers (TIA) to read the output of 2-electrode EC sensors. When used in stand-alone data-logging mode, data sampling time is arbitrary and is set from a Graphical User Interface (GUI) when the unit is connected to a PC, which allows flexibility in choosing the desired temporal resolution for monitoring the various parameters during a breathing cycle. The collected data is saved in a file on the 4Mb Electrically Erasable Programmable Read Only Memory (EEPROM), and each file is stamped with the time and date of recording acquired from the Real Time Clock Calendar (RTCC). The system communicates with an external processing and display system, e.g., a PC via a detachable plug through a USB port, or via any standard wireless data protocol such as Bluetooth, LoRaWAN, Wi-Fi, etc. A GUI displays either downloaded files or can display data being collected in real-time on a monitor. Data can also be saved on the local processing and display unit in a format chosen by the user such as the Comma Separated Value (CSV) file format. The unit can be battery operated or powered via a wall adapter.

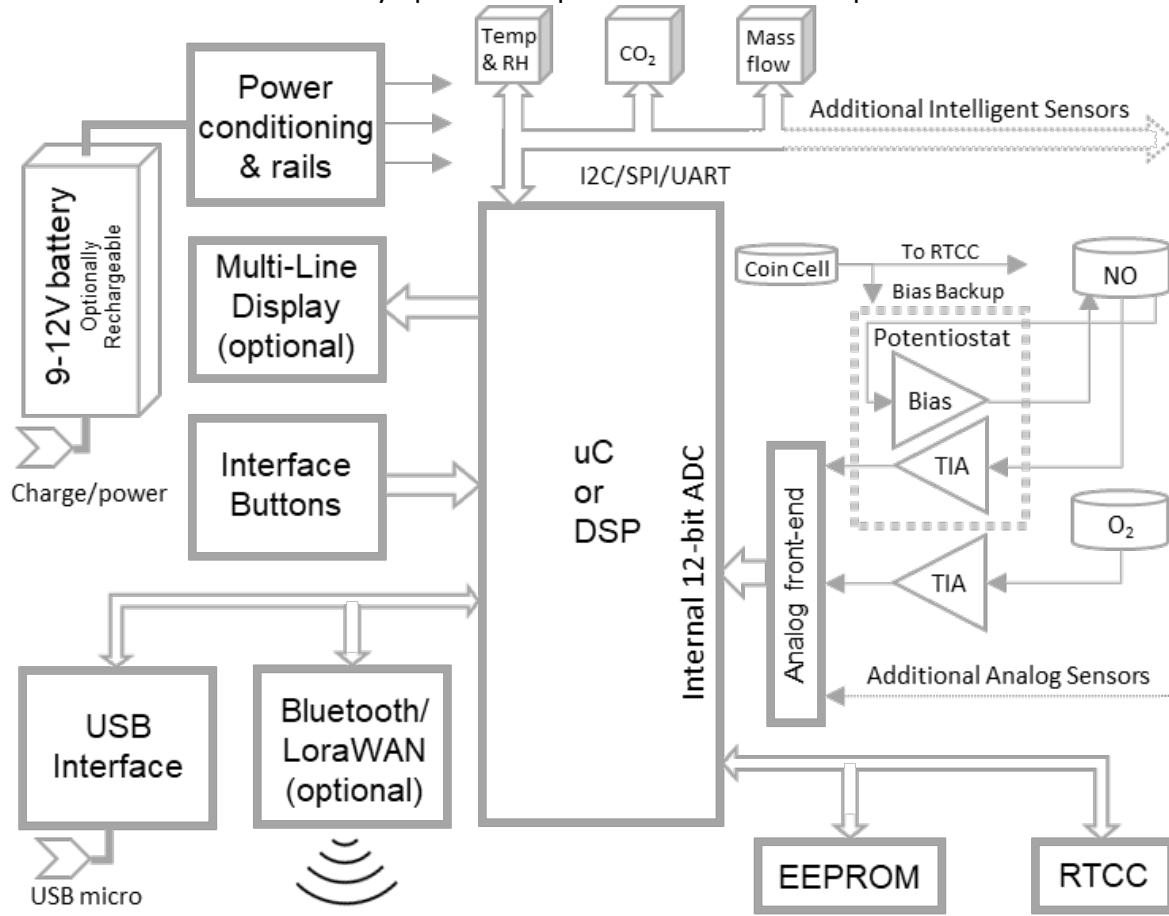

Figure 1: DLSIB Block Diagram

The prototype system, for versatility, uses terminal blocks to allow easy and rapid removal/addition of either analog or intelligent sensors as called for in the specific application. When utilized as a stand-alone data-logging unit, the system's firmware conditions and digitizes outputs from the attached analog sensors and requests binary data from the connected intelligent sensors at exact time intervals initiated by the microcontrollers built-in timers. This data, minus superfluous values such as Cyclic Redundancy Checks (CRC) from the intelligent sensors, is collected into a field which is subsequently written to the system EEPROM. The time and date of a recording along with a number of these fields comprise a complete file. The EEPROM contains a file vector table pointing to each file's starting address, and a block of control registers containing information about the installed sensors, current measurement timing settings, and system status. If any errors are encountered such as a Not Acknowledgement (NAK) from the I2C bus, recording is halted, an LED flashes to inform the user, and the error is written to the STATUS0 register in the EEPROM. When the system is controlled by the GUI via the USB port, the data for each particular sensor is requested by the GUI, displayed on a graph and recorded to local memory on the PC for analysis or recording to a local hard drive.

### B. Attached Sensors

The DLSIB system can read and log data from both analog output sensors and intelligent sensors. Intelligent sensors are devices that digitize and process analog signals from transducers. This digital representation of the quantity being measured is then stored in memory and available when requested by a processor via one of several digital protocols. These sensors are factory calibrated, and accuracy, precision and resolution specifications are usually given in the device's accompanying literature. Analog sensors produce a voltage or current (current for the EC sensors used in this work), and the end user is responsible for measuring and calibrating the output signal to produce a readout representing the actual quantity. Table I shows the sensors used with the initial prototype for equine testing, and specifications where available.

Table I: Attached Sensors and Specifications

| Device | Parameter | Min | Max | Res. | Accuracy | Type | τ |
|---|---|---|---|---|---|---|---|
| SHT31-DIS | Temperature | -40 °C | 125 °C | .01 °C | ± 0.2 °C | Digital $I^2C$ | > 2 s |
|  | RH | 0% | 100% | .01% | ± 2% | Digital $I^2C$ | 8.0 s |
| SGX-4OX-EL | $O_2$ | 0% v/v | 25% v/v | N/A | N/A | Analog | < 10 s |
| SGX-4NO-250 | NO | 0 ppm | 250 ppm | 0.5 ppm | N/A | Analog | < 40 s |
| MH-Z16 | $CO_2$ | 0 ppm | 50,000 ppm | N/A | ±50 ppm + 5% | Digital UART | < 30 s |
| SFM3000 | Mass flow | -200 SLM | +200 SLM | N/A | ± 0.1 SLM | Digital $I^2C$ | 0.5 ms |

### C. Sensor Housing

Figure 2 shows a block diagram of the physical arrangement of the sensors and the housing. Gas enters through a one-way check valve, passes through moisture exchange tubing, enters the insulated housing containing the main gas sensors, and exits via another one-way check valve followed by the mass flow sensor, which in this iteration is used to verify flow through the sensor housing.

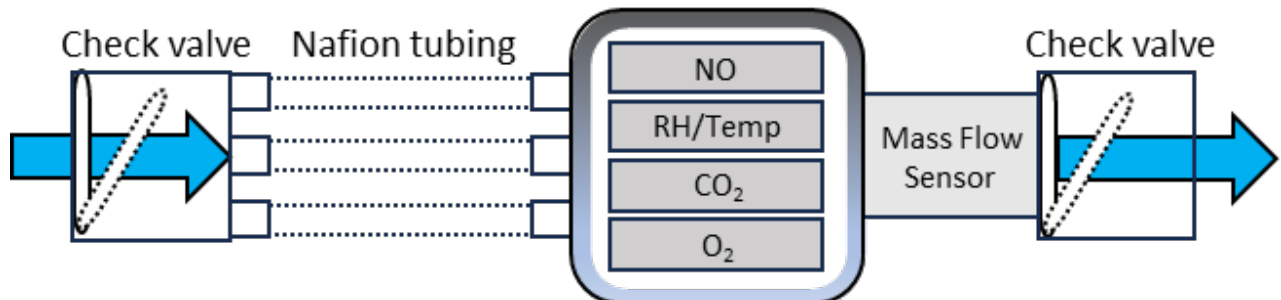

Figure 2: Physical Layout of Sensors

The moisture exchange tubing (Nafion branded), as will be seen in section III, is used to regulate the humidity of the flowing gas with ambient humidity levels. This also somewhat regulates the gas temperature with ambient conditions as well. The arrangement allows a volume of gas to flow into the housing and contains it there until the next pulse (exhalation) of gas arrives. For the initial work performed with horses by the Texas Tech University School of Veterinary Medicine, steady-state levels of exhaled gases are being measured. The veterinarian researchers will compare these recorded respiratory biomarkers with laboratory results from bronchoalveolar lavage (BAL) tests in an effort to corelate specific levels of each gas with respiratory ailments identified by the laboratory BAL results. The results of this study will be published in a separate article.

The housing and the mask that is used to collect sample gas from the horses are shown in figure 3. The mask uses several check valves to allow inhalation from ambient and to steer exhalation to tubing leading to the Nafion lines and the DLSIB housing. Nine medical-grade Nafion tubes are connected in parallel to allow easier gas flow as the inner diameter of the tubes is on the order of 1/16 inch. This also gives a larger equivalent surface area for moisture exchange to take place with ambient air.

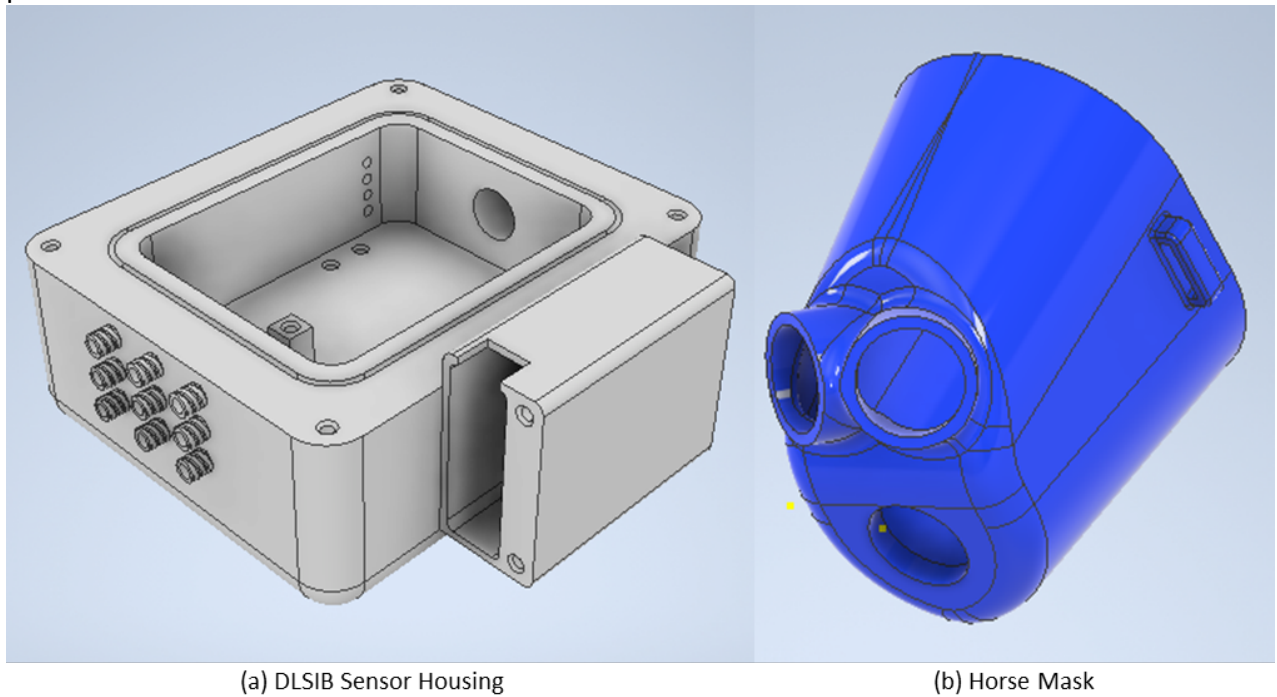

(a) DLSIB Sensor Housing          (b) Horse Mask

Figure 3: DLSIB Housing and Horse Mask

The first iteration of the DLSIB system intended for equine studies had the DLSIB housing mounted directly on the horse mask. This proved problematic as the animals often throw their heads violently in an attempt to shake the mask off. In the latest design, the DLSIB housing sits on a table and receives gas from tubing from the mask.

**D. System Software**

      Two custom programs were written for use with the DLSIB system, the first program is a user interface specifically designed to configure the DLSIB, including setting: measurement timing, installed sensors (add/delete), date and time, error checking/clearing, file deletion, etc. The interface also uploads recorded files and allows real-time data logging via the USB connection. The interface loads a calibration file when it boots up that contains amplifier correction factors for the analog front-end and calibration coefficients for each attached EC sensor. Once a DLSIB unit is connected, the interface displays the unit's serial number & firmware version, the unit's time and date, battery voltage, the installed sensors, the current

sample rate, error status and all locally recorded files. Once a file is loaded or as real-time, USB logging progresses, parameters are chosen from those available to graph in strip chart format. Data from the attached EC sensors can be viewed as raw outputs, or as corrected values using the calibration coefficients. Data can be saved on the host PC to a Comma Separated Value (CSV) file. Figure 4 shows the interface.

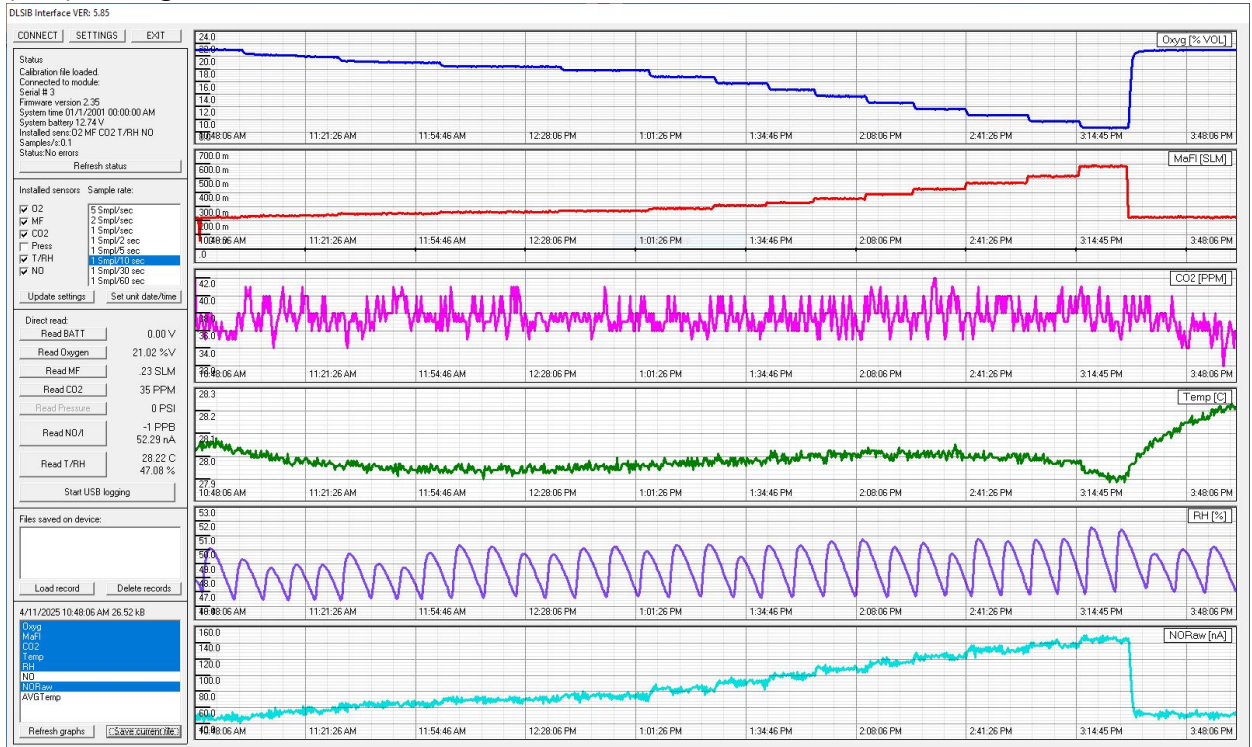

Figure 4: DLSIB Interface Software

The second program is a data set viewing and manipulation software suite with specialized calibration functions. This software allows the user to graphically select time intervals over which two data sets are averaged to create a correlated set of data points. The software then automatically uses linear regression techniques to generate coefficients for a best-fit curve of linear or polynomial type (up to 4$^{th}$ order). The software has been shown to reduce hours of work using spreadsheets for calibrations to a matter of minutes. Figure 5 displays the software being used to generate a best-fit function for an EC NO sensor's baseline output current as a function of temperature.

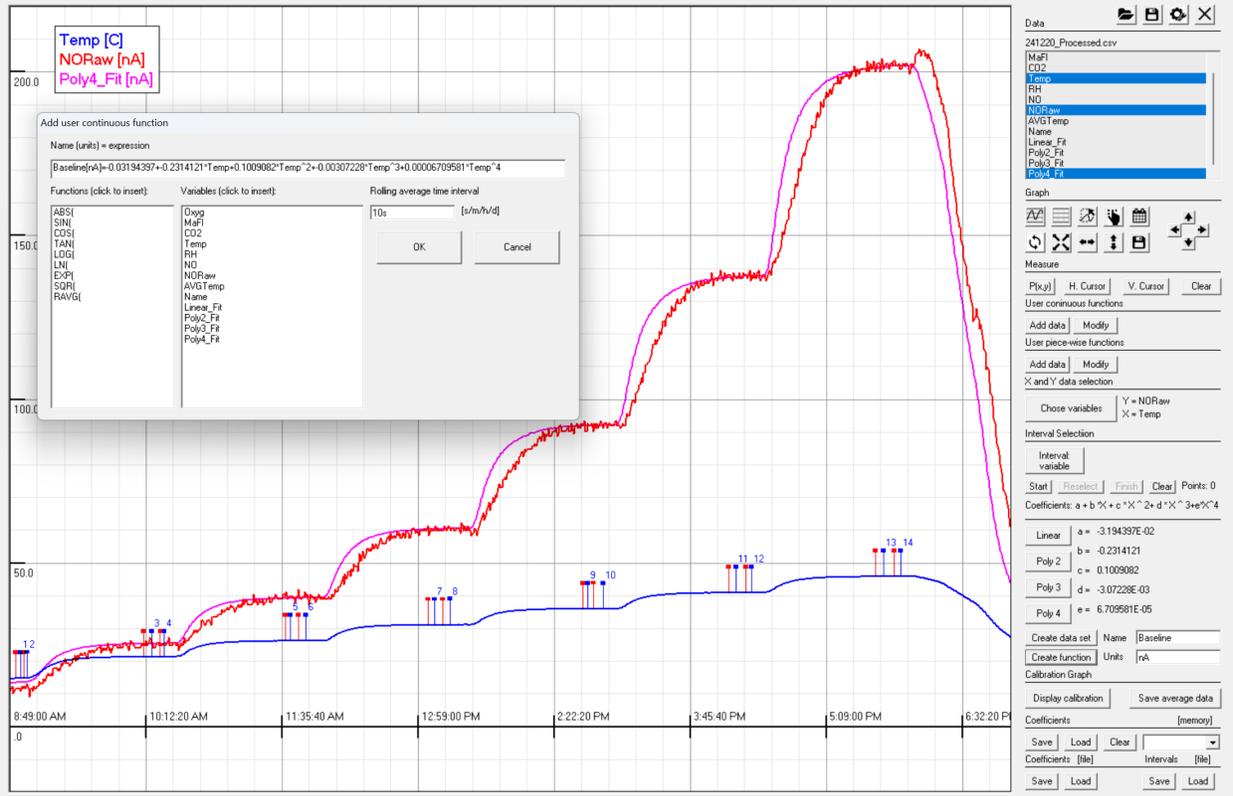

Figure 5: Calibration Utility

### III. Gas Sensor Calibration and Verification

Two types of gas sensors are utilized in the prototype: An intelligent Non-Dispersive Infrared (NDIR) type for the $CO_2$ gas sensor, and EC types for the $O_2$ and NO gas sensors. The same procedure was used for verification of both types of sensors response to its target gas, and also for calibration of the EC type sensors. Figure 6 shows the setup for gas mixing and environmental control used to calibrate and verify measurements. Each calibration bottle is connected to a Mass Flow Controller (MFC), and then the flows are combined to obtain a desired mixture.

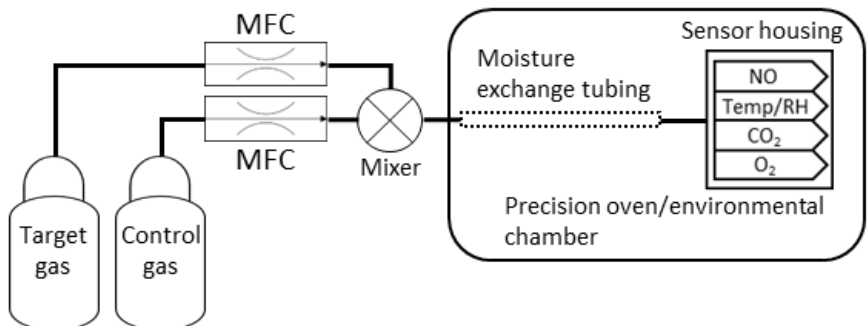

Figure 6: Gas Calibration Set Up

For calibration and validation procedures, a calibrated standard of the target gas (given concentration in ppb or ppm) meant to be picked up by the particular sensor being evaluated is precisely mixed with an inert control gas that the sensor does not respond to. The amount of the final mixture of the calibration gas is given by equation 1.

$$Target = \frac{Target\_flow}{Target\_flow + Control\_flow} * Target\_conc \quad (1)$$

As stated in section II.E, the moisture exchange tubing is used to equilibrate the humidity of the gas flowing through it with the ambient humidity level, in effect acting to regulate the flowing gases' humidity. This is necessary to keep the NDIR $CO_2$ sensor from becoming saturated with moisture and must be used to mitigate rapid fluctuations in gas humidity (on the order of 10% or more in a matter of seconds) as this is extremely detrimental to the performance of the NO sensor at its extreme low end of resolution. The sensor housing and moisture exchange tubing are placed in an environmental chamber to control temperature and steady-state RH as these factors affect the sensors and to some small degree the amplification stages of the electronics.

**A. $CO_2$ Sensor**

The $CO_2$ sensor chosen for this initial prototype was an MH-Z16 manufactured by Zhengzhou Winsen Electronics Technology Co., Ltd. This sensor is an intelligent NDIR module. It measures from 0 to 50,000 ppm (0 to 5% v/v) with a stated accuracy of ±50 ppm + 5% of the reading. This was chosen for its high range, as this is roughly at the value expected from human/equine exhaled $CO_2$ levels. While the sensor is factory calibrated, the authors thought it prudent to compare the units measured $CO_2$ values in-situ inside the DLSIB housing. This is to check for any effects on accuracy due to the passage of the gas through the moisture exchange tubing and any possible cross-sensitivity or interference from the other target gases.

Figure 7 (a) shows applied CO2 steps ranging from ~5 k ppm to ~50 k ppm. The steps are not exact due to the resolution of the Alicat flow controllers used to mix the CO2 standard and the Ultra-Pure air standard gases.

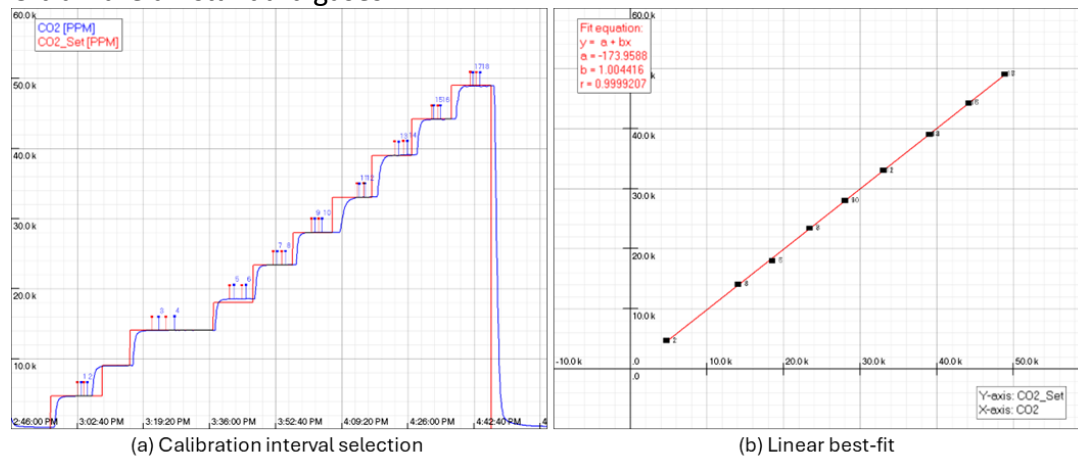

(a) Calibration interval selection     (b) Linear best-fit

Figure 7: MH-Z16 readout and CO2 Steps

A linear fit line calibration was applied to the two data sets – figure 7 (b), with a resulting R-value of 0.9999207. There was a single outlier to the data at a set-point of 18 k ppm, with the MH-Z16 reading ~ 18.54 k ppm. This equates to an error of ~2.97% (figure 8), well within the specifications of the sensor. It should be noted for the sake of completeness, however, that the researchers did observe some discrepancies in flow rates from the Alicat flow controllers over the course of the gas calibrations. As a redundant check, it was decided to add a separate flow meter in-line with the flow controllers to alert the researchers of issues.

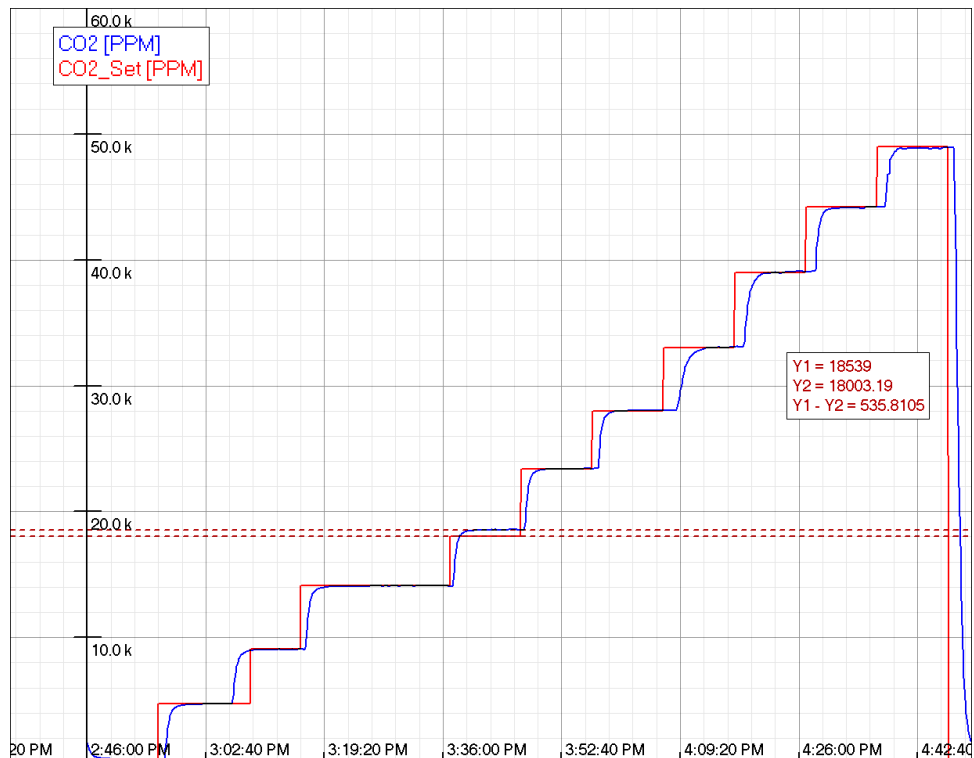
Figure 8: Maximum Error Outlier

**B. O₂ Sensor**

    The $O_2$ sensor is an EC two-terminal unit manufactured by SGX Sensortech; the model SGX-4OX-EL. The sensor generates a small electrical current that is linearly proportional to the amount of oxygen impinging on it. The range of the unit is 0 to 25 % v/v, and the stated sensitivity is 0.07 ± 0.02 mA for a full-scale reading. The sensor is designed to deliver current to a specified resistive load of 100 Ω. Practically, this is accomplished with a Transimpedance amplifier (TIA). As with all EC sensors, the sensitivity is only roughly specified, the end user must perform a calibration with each unit, even those from the same lot.

As the oxygen sensor is used well within its intended range, a very straightforward calibration method was used consisting of mixing an Ultra-Pure standard containing 21.6 % v/v of $O_2$ with varying amounts of the NO standard which contains mostly Nitrogen. Flow rates were adjusted to generate $O_2$ levels from ~9.5 to 21.5 % v/v. Figure 9 (a) displays calibration intervals being selected from the raw sensor output as well as the $O_2$ levels applied.

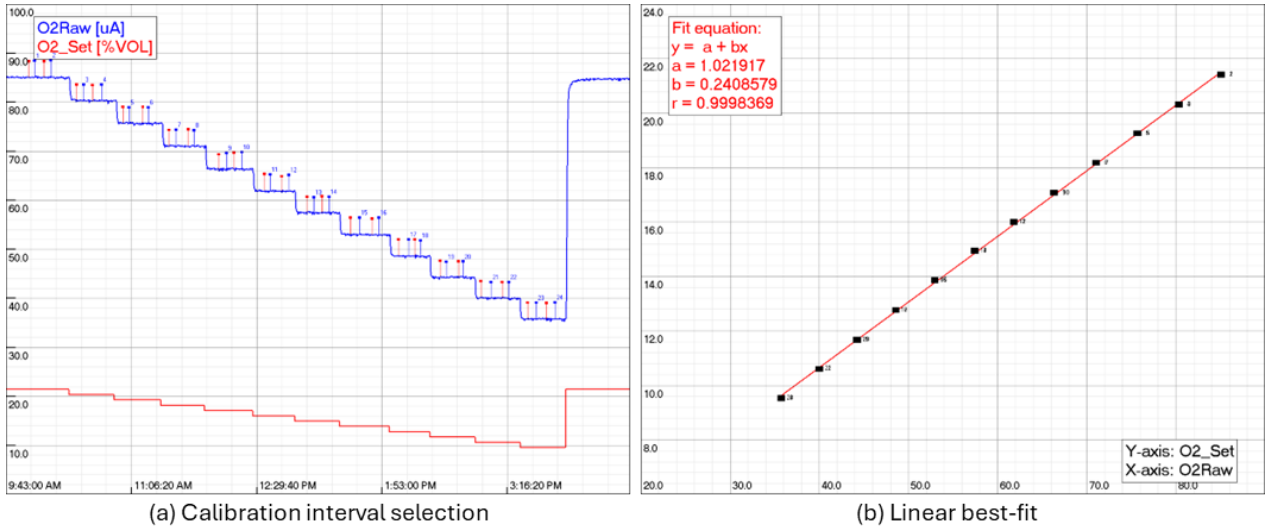

(a) Calibration interval selection

(b) Linear best-fit

Figure 9: EC O$_2$ Sensor Output, Applied O$_2$ and Calibration Intervals

Again, the calibration software was used to generate a best-fit linear function between the sensor's output and the applied O$_2$ levels. Figure 9 (b) shows the best-fit calibration, with an R-value of 0.9998369.

As stated earlier, the coefficients of this best-fit line are saved to a calibration file that the DLSIB interface loads upon boot up and uses to generate calculated O$_2$ values from raw sensor data. Figure 10 displays the applied versus calculated O$_2$ levels for this calibration.

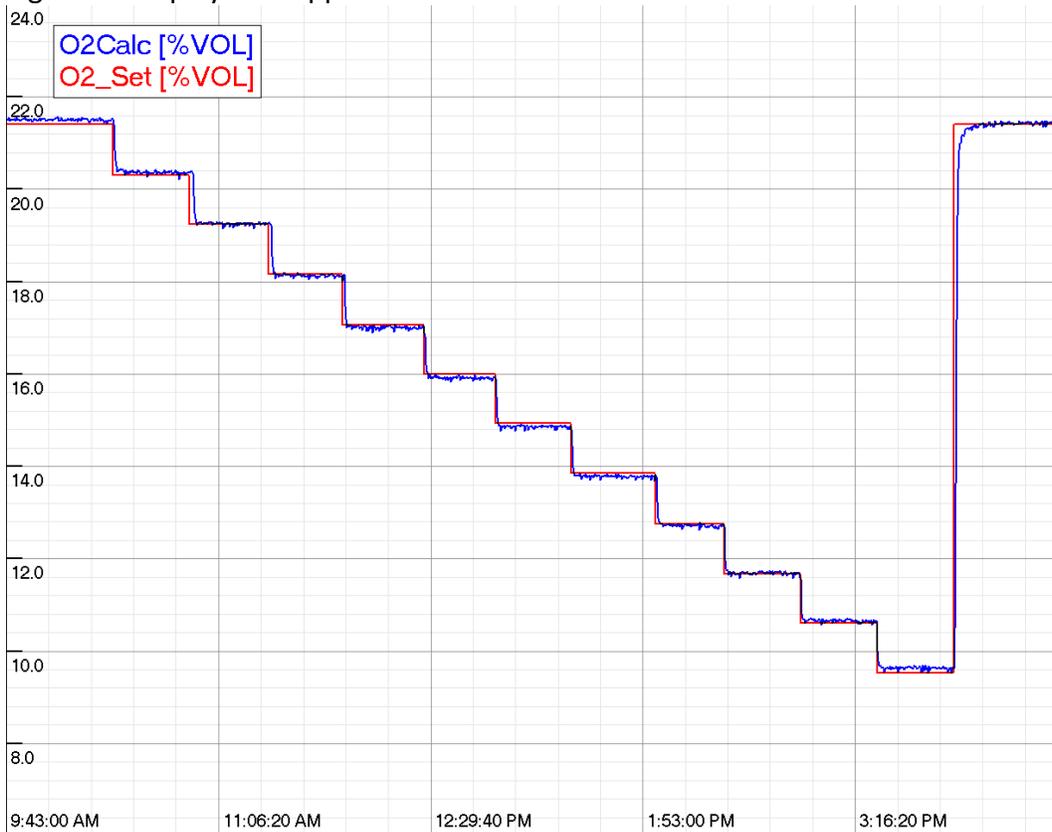

Figure 10: O2 Set and Calculated Values

The maximum deviation between the applied and calculated values is during the beginning of the calibration and equates to roughly a 0.61 % error.

**C. NO Sensor**

The NO sensor used in this work is an EC three-terminal unit, also manufactured by SGX Sensortech; the model SGX-4NO-250. The nominal range of the sensor is 0 to 250 ppm, with a stated sensitivity of 400 ± 100 nA / ppm. Three terminal EC sensors require precision, regulated bias voltages and produce a nano to micro amp electrical current based on the reaction of the target analyte (NO, in this case). This combination of bias and electric current measurement is implemented by a circuit called a potentiostat, more information about these circuits can be found here [11]. There are two components that contribute to this current signal; a baseline current that is produced when no analyte (NO) is present, and a theoretically linear current above this baseline that is proportional to the amount of the analyte present. When such a sensor is used within its nominal range, the baseline current has very little overall effect on the measurement, this is the case with the $O_2$ sensor used in this project. When attempting to use these types of sensors at ranges below their nominal levels, however, changes in shifting baseline currents must be corrected for. Both of the components of the output signal current are functions of temperature to varying degrees. For the NO EC cells used, sensitivity is a relatively weak function of temperature, while baseline current has a strong dependance on ambient temperature. This secondary effect of temperature on sensitivity compared to baseline effects has been reported elsewhere in the literature [12]. Figure 11, taken from the data sheet of the SGX-4NO-250 sensor [13], shows the variations of the sensors sensitivity to temperature (a) and baseline to temperature (b) for several samples.

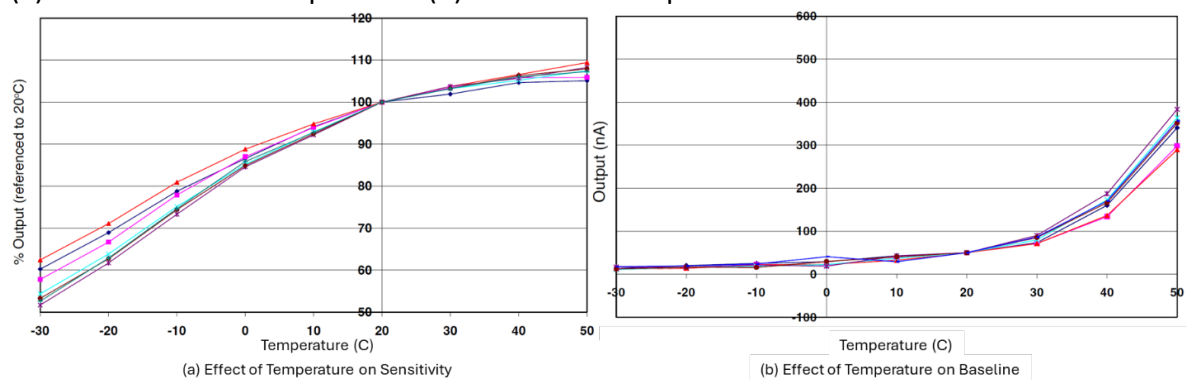

Figure 11: Temperature Dependance of Sensitivity and Baseline Signals of SGX-4NO-250 [13]

As the goal of the work was to use the NO sensors for trace gas measurements, the temperature dependance of the baseline had to be quantified and corrected mathematically as these temperature-induced variations would completely swamp the readings in the low ppb range otherwise. This large shift in baseline current compared to the low current levels produced by ppb range gas levels has been seen by researchers attempting to utilize EC NO sensors for environmental measurements [14]. Sensitivity dependance on temperature was deemed not significant enough over the temperature range of interest for this work to compensate for. Steady-state humidity was also found to cause very little change in the output signal; this has also been observed by other researchers using similar EC NO sensors [15]. Humidity transients for this project, as they will be due to expelled breath, can be quite large, often exceeding a change of 70% RH within a matter of seconds. Rapid RH changes are therefore mitigated through the use of the moisture-permeable tubing to equilibrate the RH level of the measured gas with ambient RH levels before being applied to the sensor. This was determined experimentally to be necessary as large transients cause sensor instability that can last for hours.

### 1. Mathematical Correction of Baseline-Temperature Effects and Sensitivity Modeling

The basic procedure for calibration with temperature compensation is to place the EC NO sensor and a combination temperature and humidity sensor along with the associated electronics in a precision oven and step the temperature over a range with pure air flowing over the sensor while humidity is kept constant. The interface discussed in section II.D then records the temperature, relative humidity and output signal of the sensor. The calibration program (also discussed in II.D) is then used to create best-fit coefficients from the temperature data and the EC NO sensor output signal data. These coefficients are then used to create a baseline that is a function of temperature (equation 2).

$$Baseline(T) = a_0 + a_1 T^1 + a_2 T^2 \ldots a_n T^n \, [A] \quad (2)$$

After this procedure, the temperature is held at a steady-state value, and mixtures of a NIST-traceable NO reference gas is mixed with pure air to obtain a gas with varying levels of NO. Once this data set is acquired, an excess signal function is created in the software comprised of the sensor output minus the baseline(T) signal (equation 3).

$$Exc(i, T) = (i - Baseline(T)) \, [A] \quad (3)$$

Another set of coefficients are then generated for a best-fit line between the applied NO and this excess signal (equation 4). This new data set then gives the final calculated NO in ppb as a function of both the EC sensor output signal (generally in nA) and the recorded temperature.

$$NO(i, T) = a_0 + a_1 Exc(i, T)^1 + a_2 Exc(i, T)^2 \ldots a_n Exc(i, T)^n \, [PPB] \quad (4)$$

### 2. Baseline and Sensitivity Calibration and Results

Figure 12 displays a baseline calibration from an SGX-4NO-250 EC NO sensor. In figure 12 (a), intervals are selected from the temperature curve in regions where the data reaches a roughly steady-state value, and figure 12 (b) shows the averaged data points from the NO sensors raw current output and the temperature over these time intervals, along with the auto-generated coefficients and resultant fit curve. A $4^{th}$-order polynomial was found to give a good R-value for baseline EC NO output current as a function of temperature (Eq. 2).

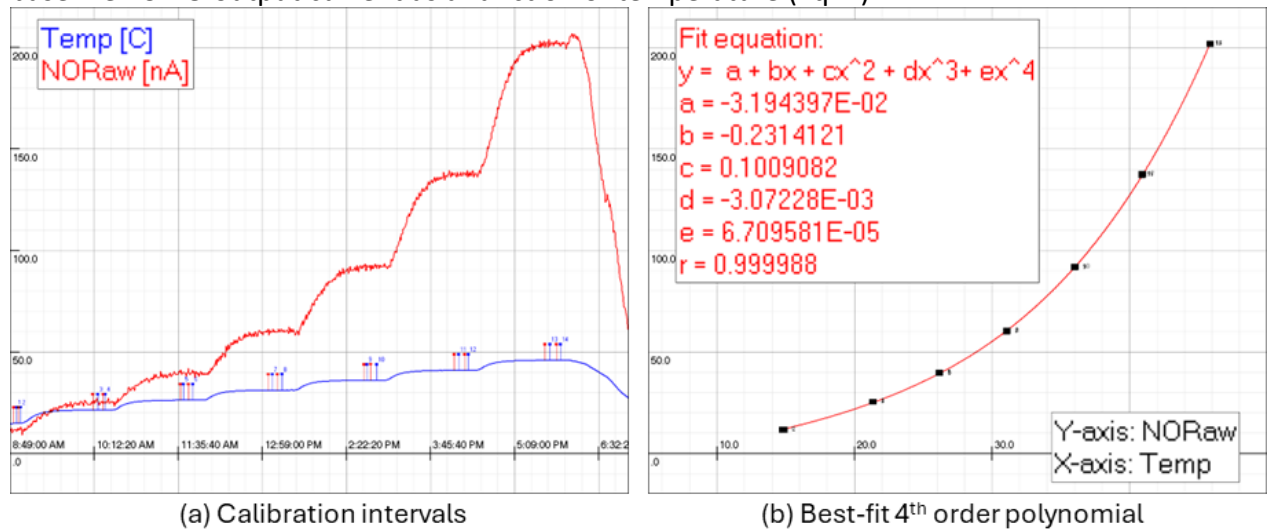

(a) Calibration intervals  (b) Best-fit $4^{th}$ order polynomial

Figure 12: Baseline Interval Selection & Coefficient Generation

Figure 13 shows the procedure for generating coefficients for the sensitivity of the same SGX-4NO-250 EC NO sensor. In this case, the two data sets of interest are the excess output current above baseline, Eq. 2, and the applied NO in ppb. Sensitivity has been verified in this work to mostly be a linear function for the EC NO sensors used thus far (Eq. 4).

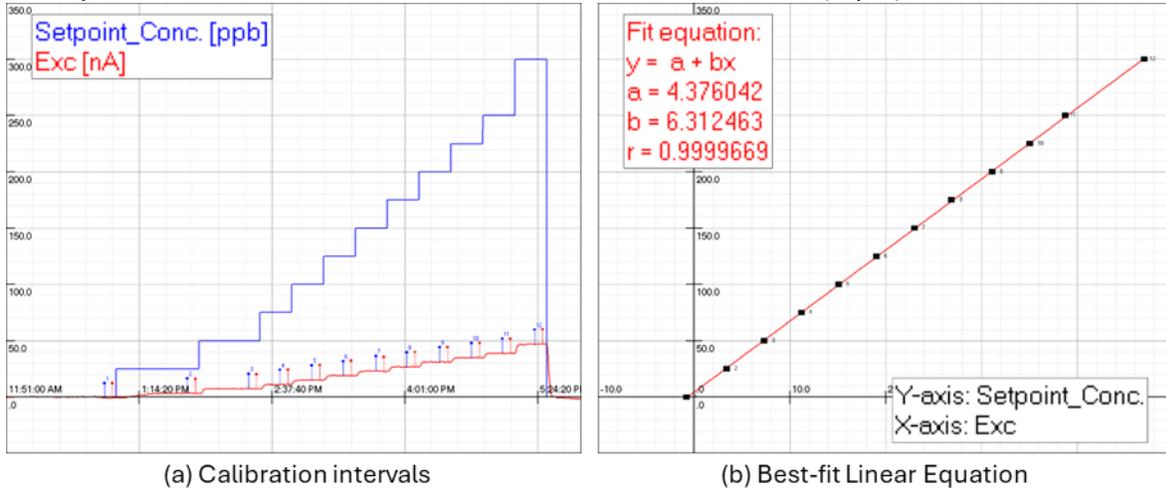

(a) Calibration intervals    (b) Best-fit Linear Equation

Figure 13: Sensitivity Interval Selection & Coefficient Generation

After coefficients are generated, they are saved to a CSV file. These coefficients are also placed in the calibration file that is loaded by the interface program. The program then, in addition to the data set containing the raw current output from the NO sensor, generates a data set using the raw output and temperature data from the SHT31-DIS sensor with a mathematically corrected value of NO in ppb. Figure 14 shows NO steps ranging from 25 ppb to 550 ppb applied to the DLSIB (red trace) and the system's corrected NO measurement (blue trace).

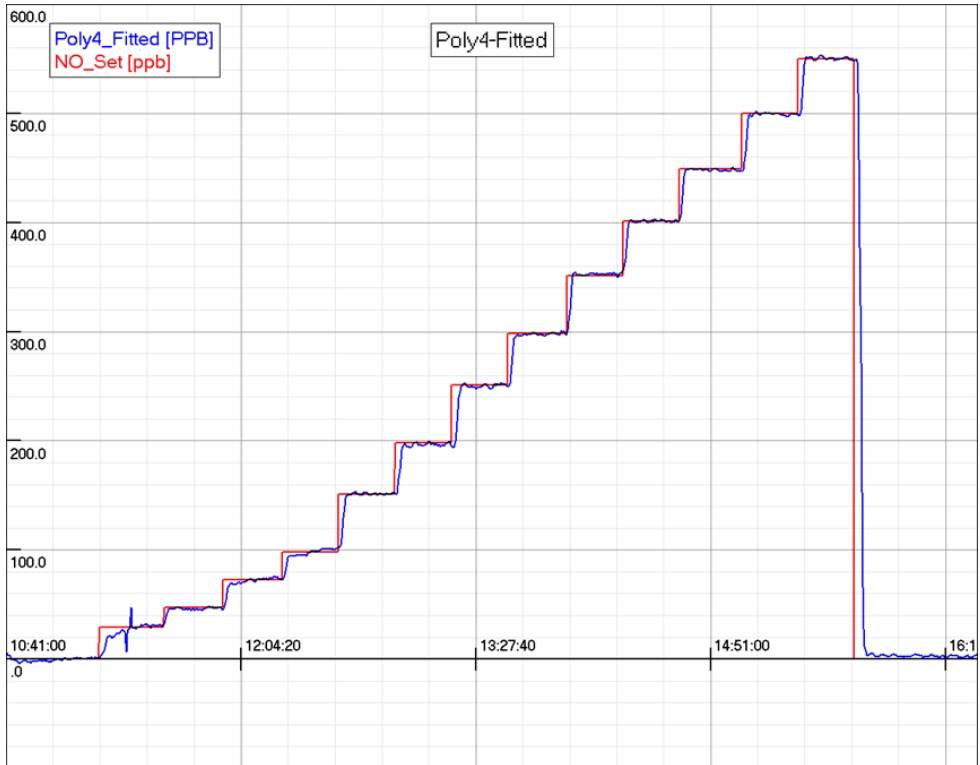

Figure 14: Applied and Calculated NO Levels

Any number of units can be used as the calibration file associates best-fit coefficients with a DLSIB unit's serial number.

## IV. DLSIB Evaluation

Following the systematic characterization of each of the sensors individual responses described in the preceding sections, the DLSIB unit was subjected to varying mixtures of all three of the gases this particular unit was designed to measure simultaneously. A slightly modified version of equation 1 was used to calculate the concentrations of each gas from the flow rates. Table II shows the flow rates, along with the calculated and measured concentrations of all three gases in this whole system evaluation.

Table II: Three gas mixture

| Time | Quantity | O$_2$ | CO$_2$ | NO |
|---|---|---|---|---|
| 9:53 AM | Flow | 0.5 SLM | 0 SLM | 0 SLM |
| | Conc | 21.6 %v/v | 0.00 kppm | 0 ppb |
| | Meas | 21.6 %v/v | 0.40 kppm | 0 ppb |
| 10:00 AM | Flow | 0.5 SLM | 0.36 SLM | 0 SLM |
| | Conc | 21.1 %v/v | 20.9 kppm | 0 ppb |
| | Meas | 21.2 %v/v | 21.0 kppm | 0 ppb |
| 10:20 AM | Flow | 0.5 SLM | 0.36 SLM | 0.127 SLM |
| | Conc | 18.4 %v/v | 18.2 kppm | 127 ppb |
| | Meas | 18.3 %v/v | 18.6 kppm | 132 ppb |
| 10:40 AM | Flow | 0.5 SLM | 0 SLM | 0.127 SLM |
| | Conc | 17.2 %v/v | 0.00 kppm | 200 ppb |
| | Meas | 17.0 %v/v | 0.40 kppm | 202 ppb |
| 11:00 AM | Flow | 0.5 SLM | 0 SLM | 0 SLM |
| | Conc | 21.6 %v/v | 0.00 kppm | 0 ppb |
| | Meas | 21.6 %v/v | 0.40 kppm | 0 ppb |

Figure 15 is a graph of the values calculated and recorded by the DLSIB system for this testing procedure presented in separate, strip-chart form. It should be noted that the mass flow graph is the measured mass-flow exiting the DLSIB housing. This was done for initial testing with animals for verification of flow as check valves must be employed to steer inhalation and exhaust.

The initial drop in the NO sensor output upon initiation of gas flow is a common occurrence, and the authors see this frequently. It is suspected that this results from a combination of fast RH changes and a difference between the temperature of the impinging gas and the sensor.

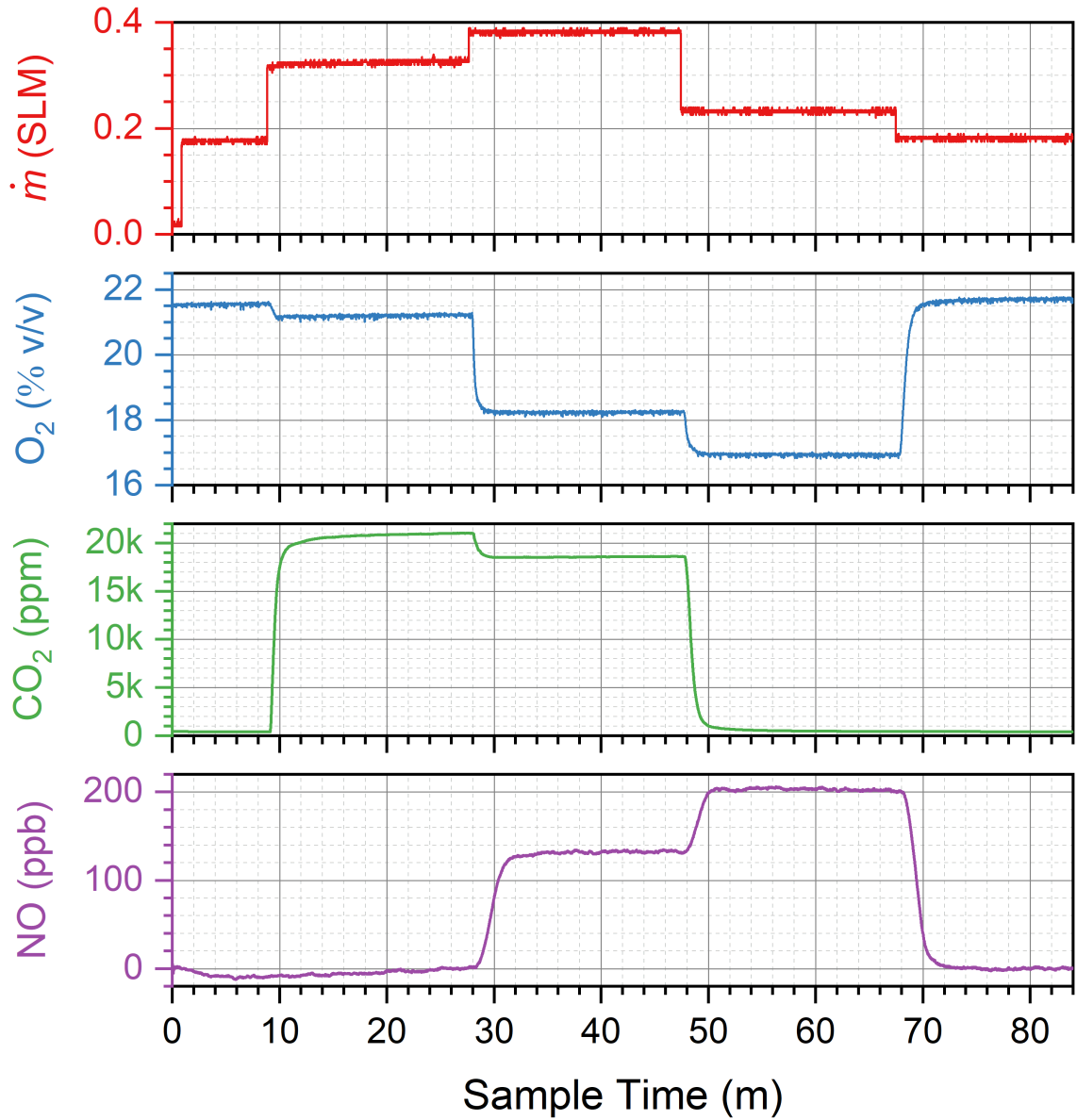

Figure 15: Three Gas Blend Testing
This multiple gas testing was performed without controlling either the ambient temperature or the relative humidity. During the testing, the temperature (as monitored inside the insulated DLSIB housing) rose from approximately 25.3°C to 29.2°C. As stated in section III.C, the NO sensor baseline output is a very strong function of temperature, and this must be compensated for mathematically. Figure 16 displays the raw NO current output during the course of this test, the temperature and the calculated, temperature-compensated baseline. From the graph, it can be seen that the calculated baseline tracks the sensor's actual baseline output very nicely.

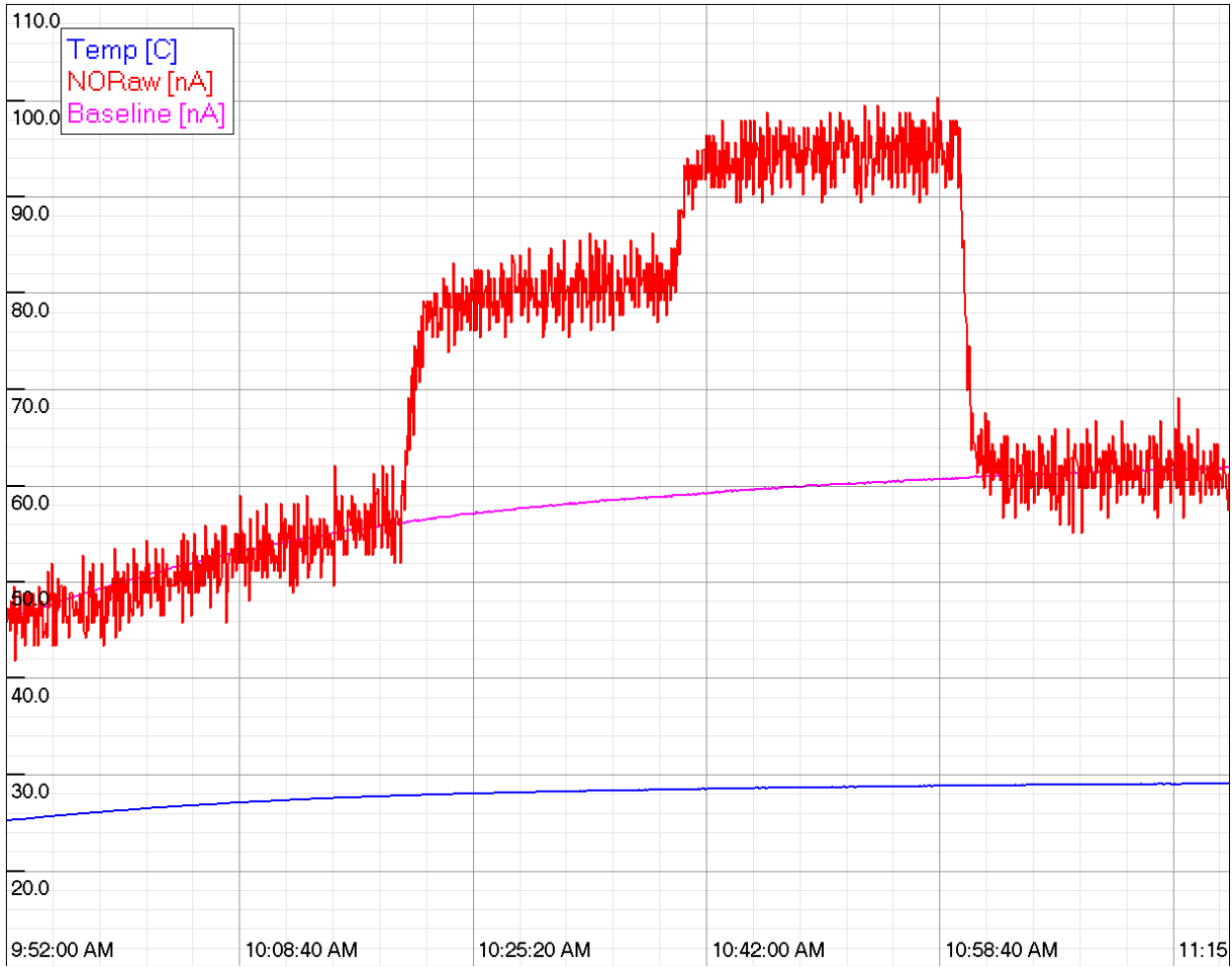

Figure 16: Raw NO Sensor Output and Calculated Baseline.

Figure 17 shows the measured NO (with a rolling average applied for clarity) using both the calculated baseline as well as a fixed constant baseline for the purpose of illustrating the efficacy of this temperature compensation: without correcting for temperature, an approximate change of 4°C corresponds to a change in NO reading of roughly 90 ppb.

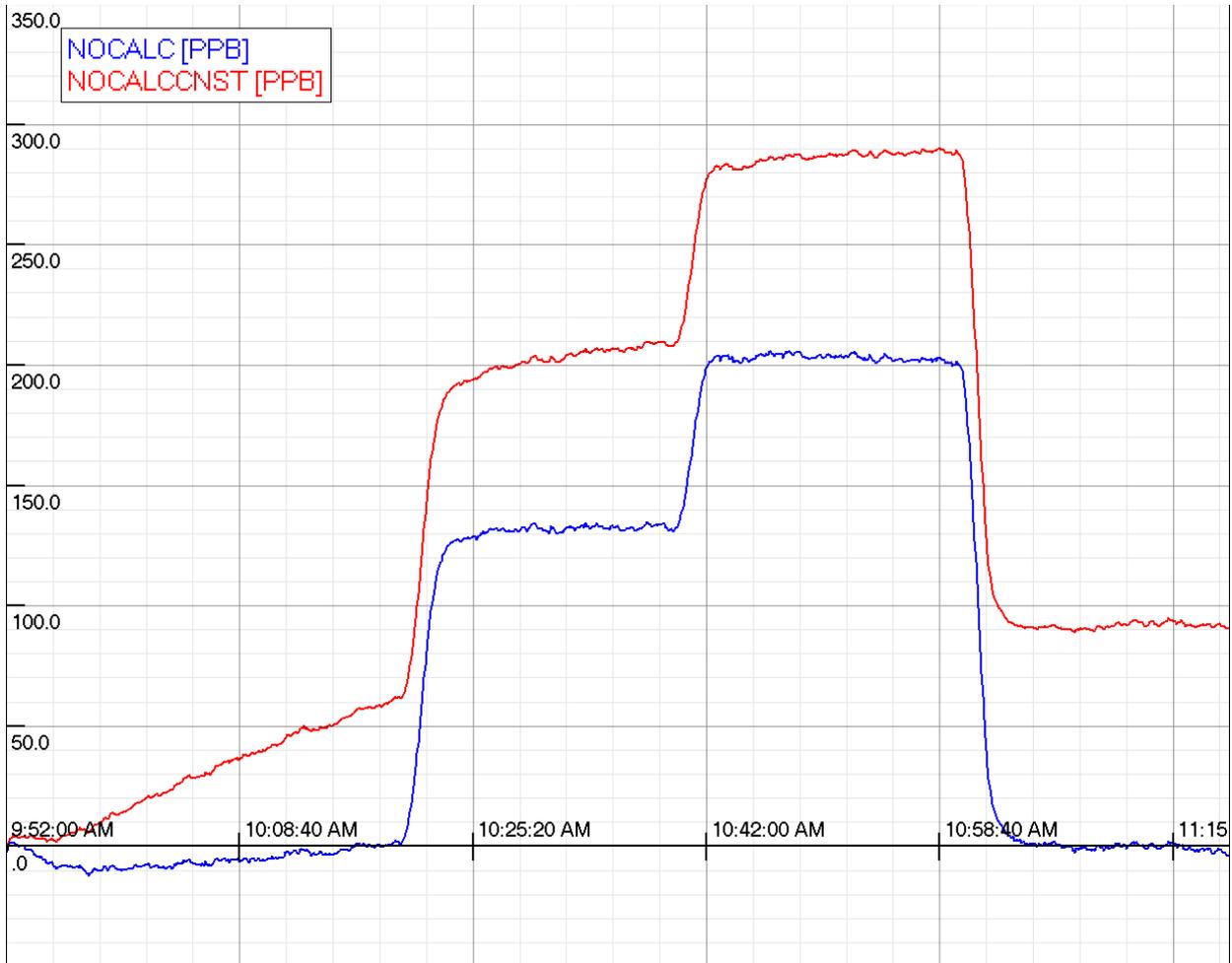

Figure 17: Calculated NO using both Constant and Compensated Baselines.

**V. Future Work**

The core DLSIB system, configured for the three main gases initially of interest, has been successfully designed and characterized. Ongoing work will revolve around tailoring the system specifically for use with human and equine exhaled breath. The authors have already determined that there is a component of human exhalent that the NO sensor is inversely cross-sensitive to. Research of prior literature [16,17] points to this most likely being either Carbon Monoxide (CO) or Ammonia (NH3). Testing with these gases using various filtering approaches is currently underway.

Additionally, while the system is accurate enough for the initial specifications laid out, the authors would like to research the effects of temperature on sensitivity, as well as the effects of the absolute value of humidity (and to a much lesser degree, pressure) on both baseline and sensitivity. These secondary and tertiary effects cause minute shifts in baseline (the effects are on the order of a few nanoamps and are zeroed out during the beginning of measurements) and also on the sensitivity.

**VI. Conclusion**

A portable data logging system was designed with both analog inputs, as well as SPI, UART and I2C digital communication channels for interfacing with modern intelligent sensors. The system has built-in, specialized low-noise, high gain potentiostat and TIA circuitry to interface directly with EC gas sensors. A custom program was written for the unit to control settings, upload recorded files or record from the unit in real-time via a USB connection. The interface graphs the parameters simultaneously in separate strip charts and can save data locally in the form of CSV files. Another software utility was written to display and manipulate recorded data sets, and to perform calibrations for the various sensors. The data viewing/manipulation utility allows complete calibrations to be performed in a matter of minutes. The system utilizes moisture exchange tubing to regulate RH, and to a lesser degree temperature with ambient environmental conditions to allow for more precise and repeatable gas measurements. The system has widespread potential uses in both human and animal health care, and in scientific research focused on the respiratory system and both the immediate and long-term effects of the environmental air quality on the respiratory system.